# A Finite Element framework for computation of protein normal modes and mechanical response

Mark Bathe

*Arnold Sommerfeld Zentrum für Theoretische Physik and Center for NanoScience*
Ludwig–Maximilians–Universität München
Theresienstrasse 37, 80333 Munich, Germany

**Abstract**

A coarse-grained computational procedure based on the Finite Element Method is proposed to calculate the normal modes and mechanical response of proteins and their supramolecular assemblies. Motivated by the elastic network model, proteins are modeled as homogeneous isotropic elastic solids with volume defined by their solvent-excluded surface. The discretized Finite Element representation is obtained using a surface simplification algorithm that facilitates the generation of models of arbitrary prescribed spatial resolution. The procedure is applied to compute the normal modes of a mutant of T4 phage lysozyme and of filamentous actin, as well as the critical Euler buckling load of the latter when subject to axial compression. Results compare favorably with all-atom normal mode analysis, the Rotation Translation Blocks procedure, and experiment. The proposed methodology establishes a computational framework for the calculation of protein mechanical response that facilitates the incorporation of specific atomic-level interactions into the model, including aqueous-electrolyte-mediated electrostatic effects. The procedure is equally applicable to proteins with known atomic coordinates as it is to electron density maps of proteins, protein complexes, and supramolecular assemblies of unknown atomic structure.

**Introduction**

Equilibrium conformational fluctuations of proteins about their folded, native structure play an important role in their biological function.[1-4] Three prominent approaches used to compute conformational fluctuations of proteins are, in order of increasing computational efficiency and decreasing modeling resolution, molecular dynamics (MD), all-atom normal mode analysis (NMA), and coarse-grained elastic NMA (eNMA). MD attempts to sample the equilibrium distribution of states in the vicinity of the native structure via time-integration of Newton's equations of motion, typically modeling solvent explicitly.[5] All-atom NMA assumes harmonic fluctuations about the native state in solving the free vibration problem for the protein while treating the solvent implicitly.[2,3,6] Finally, eNMA employs a coarse-grained elastic description of the protein in which specific atomic interactions are replaced by a simple network of linear elastic springs, typically connecting Cα atoms within an arbitrary cut-off radius.[7,8] Successively coarser and thus computationally more efficient eNMA descriptions are obtained by reducing the total number of interaction sites in the system.[9-12] The idea of treating proteins as effective elastic media in calculating their normal modes dates back at least to Suezaki and Go.[13]

Despite their relative simplicity, elastic coarse-grained models have proven remarkably successful in calculating the slow, large length-scale vibrational modes of proteins and their supramolecular assemblies. As shown recently by Lu and Ma,[14] their success may partially be attributed to the fact that biomolecular *shape* plays a dominant role in determining the lowest normal modes of proteins. Indeed, large length-scale modes naturally average over heterogeneous interactions present at atomic length-scales, thereby rendering elastic descriptions valid in this regime. Global structural averages such as backbone fluctuations and inter-residue correlations are in turn also successfully predicted because they are dominated by these low frequency modes.

The success of eNMA motivates the current work, in which the elastic network model for proteins is cast in the framework of the well established Finite Element Method (FEM).[15,16] In formulating the model, the protein is defined by its mass density, $\rho$, isotropic elastic modulus, $E$, and solvent-excluded surface (SES), which is obtained by rolling a water molecule-size probe-sphere over its van der Waals surface.[17-20] As an initial exploration of the utility of the FEM in analyzing protein mechanical response, the normal modes of a mutant of T4 lysozyme and of F-actin are computed, as well as the critical Euler buckling load of F-actin when subject to axial compression. NMA results for T4 lysozyme are compared with all-atom NMA, the Rotation Translation Blocks (RTB) procedure,[21,22] which treats residues as rigid but retains atomic-level interactions as modeled by the implicit solvent force-field EEF1,[23] and experiment.

Similar to eNMA, the proposed FE-based procedure offers several advantages over all-atom NMA, including the elimination of costly energy minimization that may distort the initial protein structure, direct applicability to x-ray data of proteins with unknown atomic structure,[24-26] and a significant speed-up of the NMA itself due to a drastic reduction in the number of degrees of freedom simulated.

Additionally, the FEM offers several distinct advantages over existing elastic network models that provide the primary motivation for the current work. Principal among these is the suitability of the FEM to calculate the mechanical response of proteins and their supramolecular assemblies to applied bending, buckling, and other generalized loading scenarios, which is needed to probe the structure-function relation of supramolecular assemblies such as viral capsids,[27,28] microtubules,[29,30] F-actin bundles,[31,32] and molecular motors.[33,34] Moreover, casting the coarse-grained elastic model in the framework of the FEM opens two important avenues of model refinement that are currently being pursued. First, the atomic Hessian can be projected onto the FE-space in order to incorporate atomic-level interactions into the model, thereby eliminating the *a priori* assumption of homogeneous isotropic elastic response. This idea is similar to the initial version of the Rotation Translation Blocks (RTB) procedure proposed in Durand et al.,[35], as well as related works in modeling crystals.[36,37] The incorporation of atomic-level interactions may be particularly important in modeling binding interfaces present between constituent monomers in supramolecular assemblies such as F-actin,

MTs, and viral coat protein subunits, particularly near the onset of mechanical failure. Second, the FE-based protein model may be coupled directly to field calculations including the Poisson–Boltzmann Equation to model solvent-mediated electrostatic interactions[38-40] and the Stokes Equations to model solvent-damping in dynamic response calculations.[41,42]

**Methods**

The FEM is a mature field that is discussed in detail in references such as Bathe[15] and Zienkiewicz and Taylor.[16] Accordingly, the focus here is on its application to proteins and readers are kindly referred to the above-referenced books for details on its theoretical foundations.

Generation of the FE model requires three steps: (1) definition and discretization of the protein volume; (2) definition of the local effective mass density and constitutive behavior of the protein; and (3) application of boundary conditions such as displacement- or force-based loading. The protein volume is defined by its bounding SES, which is also called the Richards Molecular Surface or simply the Molecular Surface. This surface is defined by the closest point of contact of a solvent-sized probe-sphere that is rolled over the van der Waals surface of the protein, which defines the molecular volume that is never penetrated by any part of the solvent probe-sphere.[17-19] The SES is computed using MSMS ver. 2.6.1, which generates a high density triangulated approximation (one triangular vertex per $Å^2$) to the exact SES.[20] The MSMS-discretized SES is subsequently decimated to arbitrary prescribed spatial resolution using the surface simplification algorithm QSLIM.[43-45] The QSLIM algorithm employs iterative vertex-pair contraction together with a quadric error metric to retain a near-optimal representation of the original surface while reducing the total number of faces by an arbitrary, user-specified amount.[43] The protein volume that is bounded by the closed SES is subsequently discretized with 3D tetrahedral finite elements via automatic mesh generation using the commercial Finite Element program ADINA ver. 8.4 (Watertown, MA, USA). Application of the proposed FE-based procedure directly to x-ray data would require definition of the molecular volume from the electron density map using Voronoi tessellation or a similar procedure, as proposed by Wriggers et al.,[46], and performed by Ming et al.[24]

The protein constitutive response is modeled using the standard Hooke's law, which treats the protein as a homogeneous, isotropic, elastic continuum with Young's modulus $E$ and Poisson ratio $v$.[47] While this is conceptually similar to elastic-network based models, it is rigorously distinct: Elastic network models typically connect Cα atoms by springs of equal stiffness, which results in general in a locally *an*isotropic and *in*homogeneous elastic material with length-scale dependent mechanical properties. In contrast, the FE-model defined here treats the protein as strictly homogeneous, with an isotropic elastic material response that is length-scale invariant. The mass density of the protein is taken to be homogeneous, although it could equally be defined as a spatially-varying function from the underlying atomic constitution or from electron density data.

Finally, arbitrary boundary conditions consisting of displacement- or force-based loading may be applied to the molecule, modeling the effects of the protein environment. In the current application, the free vibration problem is solved for T4 lysozyme and F-actin in the absence of any boundary condition and the linearized buckling problem for F-actin is solved by applying co-axial compressive point loads to the ends of the molecule.

Given the protein volume, constitutive behavior, and boundary conditions, the FEM uses numerical volume-integration to derive a set of algebraic equations that is linear in the finite element nodal displacement degrees of freedom, $\mathbf{u}$,

$$\mathbf{M}\ddot{\mathbf{u}} + \mathbf{K}\mathbf{u} = \mathbf{R} \qquad (1)$$

where $\mathbf{M}$ is the diagonal mass-matrix, $\mathbf{K}$ is the elastic stiffness matrix, and $\mathbf{R}$ is a forcing vector that results from natural (force-based) boundary conditions.[15] In the case of the free vibration problem relevant to the NMA of proteins, $\mathbf{R} = 0$. Substitution of the oscillatory solution, $\mathbf{u} = \mathbf{y}\cos(\omega t + \gamma)$, into the free-vibration form of Eq. (1) results in the generalized eigenvalue problem,

$$\mathbf{K}\mathbf{y} - \omega^2 \mathbf{M}\mathbf{y} = 0 \qquad (2)$$

which after definition of the eigenvalues, $\lambda := \omega^2$, may be written in secular form,

$$\det|\mathbf{K} - \lambda\mathbf{M}| = 0. \qquad (3)$$

Various efficient FE procedures exist to obtain the solution to the generalized eigenvalue problem, yielding the eigenvalues and eigenvectors, $(\lambda_i, \mathbf{y}_i)$. In the present application an accelerated subspace-iteration method[15,48] is used for T4 lysozyme and F-actin. The substructure synthesis procedure[49] commonly available in structural mechanics FE programs could also be applied to calculate the normal modes of F-actin, as recently proposed by Ming et al.[50]

The eigenvectors corresponding to the FE nodal degrees of freedom are linearly interpolated to the Cα positions given by the atomic coordinates that were used to define the FE model. Standard equilibrium thermal averages may then be computed in the standard way, including the fluctuation of Cα atom $i$ due to mode $k$, $\langle \Delta r_{ik}^2 \rangle = k_B T a_{ik}^2 / \lambda_k m_i$, the total fluctuation of Cα atom $i$ due to all modes, $\langle \Delta r_i^2 \rangle = \sum_k \langle \Delta r_{ik}^2 \rangle$, correlations in positional fluctuations of Cα atoms $i$ and $j$, $C_{ij} = \langle \Delta r_i \cdot \Delta r_j \rangle / \left( \langle \Delta r_i \cdot \Delta r_i \rangle \langle \Delta r_j \cdot \Delta r_j \rangle \right)^{1/2}$, where $\langle \Delta r_i \cdot \Delta r_j \rangle = k_B T \sum_k a_{ik} a_{jk} / \left( \lambda_k \sqrt{m_i} \sqrt{m_j} \right)$, and the overlap, $R_{ij}$, between normal modes $i$ and $j$, defined by the inner product of the modes, $R_{ij} = \mathbf{a}_i \cdot \mathbf{a}_j / |\mathbf{a}_i||\mathbf{a}_j|$, where $(1 \leq R_{ij} \leq -1)$.[6] As with elastic network models, the protein stiffness-scale ($E$) is unknown. Accordingly, the acoustic wave speed, $\sqrt{E/\rho}$, which is the relevant physical unit in the free-vibration problem, is adjusted to best-fit the pertinent Cα fluctuation data, which is either experimental or that from the all-atom NMA. In the case of F-actin, the average mass density, $\rho$, is set explicitly and the Young's modulus is determined by matching its stretching stiffness to experiment,[51] as also performed by ben-Avraham and Tirion[52] and described in more detail below. The Poisson ratio is taken to be 0.3 for T4 lysozyme and F-actin, which is typical of crystalline solids. While the choice of $\nu = 0.3$ has, to the best of the author's knowledge, no rigorous justification, it is noted that its precise value does not affect the computed results within the range of $(0.3 \leq \nu \leq 0.5)$. This is typical of response calculations such as those performed here, in which material compressibility does not play an important role.

Two important considerations in generating the FE model are the choice of the probe-sphere radius used to define the protein volume and the degree of surface simplification performed. Regarding the choice of probe-sphere radius, two approaches were deliberated here. In the first, the probe-sphere radius is treated as an adjustable parameter, akin to the cut-off radius used in elastic network models. In this case, as the radius of the probe-sphere is increased, protein cavities in which solvent would normally be present become part of the effective elastic medium constituting the protein. Accordingly, the shape of the protein becomes a function of the probe-sphere radius, which will affect its mechanical response. In the second approach, the probe-sphere radius is treated as a fixed, physically-based parameter that is approximately equal to the size of a water molecule, as in electrostatic field calculations.[53-55] The homogeneous elastic medium of the protein is then strictly applied to those volumetric regions in which dense intramolecular packing involving close-ranged van der Waals, hydrogen-bond, and bonded interactions are present, and the molecular surface is a well-defined physical feature of the protein. The latter approach was taken here in order to retain the physical connection to atomic packing in solids.

An important theoretical property of the FEM is that it guarantees convergence to the exact solution of the underlying mathematical model as the FE mesh is refined, where the mathematical model is defined by the protein's analytical SES, constitutive behavior, and boundary conditions.[15] Thus, any normal mode or mechanical response calculation performed using the proposed FE-based procedure should in principle systematically refine the discretized representation in order to ensure convergence of the computed model property to its exact result. In practice, however, the permissible degree of surface simplification using QSLIM or similar algorithm will depend on the sensitivity of the computed observable to details of molecular shape, which must be evaluated on a case-by-case basis, as addressed below for T4 lysozyme and F-actin.

*T4 lysozyme*

The initial structure of the 164 residue (18.7 kDa) mutant T4 phage lysozyme is taken from Matsumura *et al.*,[56] (Protein Data Bank ID 3LZM).[57] CHARMM ver. 33a1[58] is used with the implicit solvation model EEF1[23] to build in coordinates missing in the crystal structure and to perform energy minimization and NMA. Steepest descent minimization followed by adopted-basis Newton–Raphson minimization is performed in the presence of successively reduced harmonic constraints on backbone atoms to achieve a final root-mean-square (RMS) energy gradient of $5\times10^{-4}$ kcal/(mol Å) with corresponding RMS deviation between the x-ray and energy-minimized structures of 1.3 Å (Fig. 1a). All-atom (ATM) and RTB NMA[21] are used as implemented in CHARMM,[22] using one-block per residue for the RTB calculations.

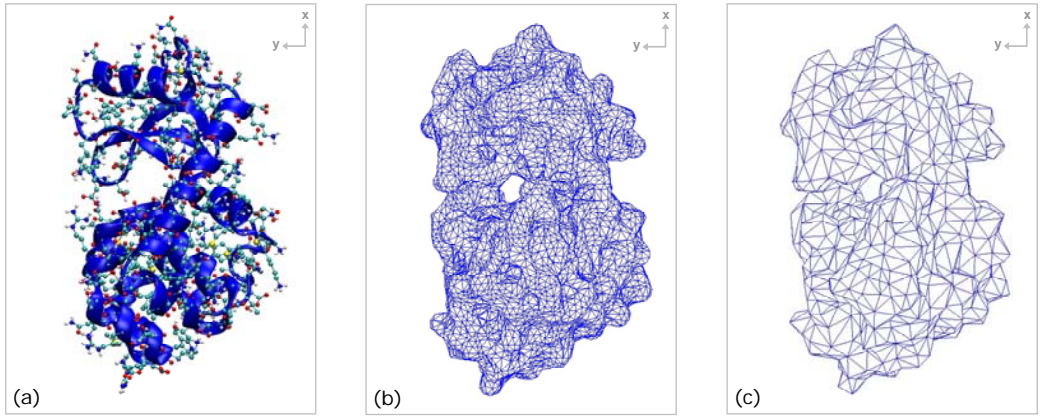

**Fig. 1** T4 lysozyme (a) crystal structure (Protein Data Bank ID 3lmz), (b) MSMS-triangulated SES, and (c) QSLIM-decimated SES used for the FE computation. Atomic structure rendered with VMD ver. 1.8.5[59] and triangulated models rendered with ADINA ver. 8.4.

To define the FE model, MSMS is used to compute the SES of the energy-minimized structure of T4 lysozyme using the MSMS-default 1.5 Å radius probe ignoring hydrogens. As noted previously, the FE model may be defined directly from the atomic structure without initial energy minimization, however, the energy-minimized structure is used here to be consistent with the ATM model, which requires minimization. MSMS generates a triangulated approximation to the analytical SES that consists of 17,300 triangular faces (Fig. 1b). This model is decimated using QSLIM to a reduced model consisting of 2,000 faces (Fig. 1c). The decimated surface-mesh is read into ADINA ver. 8.4 and used as a template to generate 6,843 4-node tetrahedral finite elements consisting of 1,627 nodes. Calculation of the 100 lowest non-rigid-body modes using an accelerated subspace-iteration method[15,48] required 27 MB of RAM and about 10 seconds on a 2.1 GHz Intel Core2Duo processor. Refining drastically the surface representation from 2,000 faces to 17,300 faces (and associated volume discretization) or computing more than 100 normal modes did not alter the Cα fluctuations significantly.

*F-actin*

The atomic structure of F-actin (52 protomers, 2.2 MDa molecular weight) is generated using FilaSitus ver. 1.4[60] based on the Holmes fiber model[61] and the structure of G-actin:ADP:$Ca^{2+}$ from the actin-gelsolin segment-1 complex.[62,63] This structure of F-actin-ADP models the filament in its "young" state when the DNase I binding region of subdomain 2 of G-actin (residues 40–48) is in its disordered loop conformation as opposed to its ordered α-helix conformation.[63-65] Importantly, in its disordered loop conformation this region forms intramolecular contacts in F-actin that stabilize the filament and have direct consequences on its mechanical properties.[66-68]

Calculation of the SES using MSMS and a 3 Å radius probe[a] results in a model with 1,248,038 triangular faces, which is subsequently decimated in several seconds using QSLIM to a reduced model with 40,000 triangular faces. The decimated surface-mesh is read into ADINA ver. 8.4 and used as a template to generate 134,883 4-node tetrahedral finite elements consisting of 31,881 nodes (Fig. 2b). Planar axial stretching is used to determine the effective Young's modulus of F-actin, $E = 2.69$ GPa, by fitting its computed value to its experimentally-measured value in the absence of tropomyosin, 43.7 nN.[51] The homogeneous mass density, $\rho = 1,170$ kg/$m^3$, is based on the 42 kDa molecular weight of G-actin and the calculated molecular volume of F-actin, which is equal to $3.1 \times 10^6$ $Å^3$ for the 52-mer considered. Normal mode analysis using the accelerated subspace-iteration procedure in ADINA requires 22 MB and less than 10 seconds to calculate the lowest 10 modes on a 2.1 GHz Intel Core2Duo processor. To test convergence of the FE solution to the exact solution, the FE mesh was coarsened considerably to a model consisting of only 7,558 4-node tetrahedral volume elements (4,000 surface triangles), for which the lowest four eigen-frequencies increased by at most 15% with respect to the more detailed model. Further mesh refinement beyond 40,000 surface elements was precluded by the problematic surface mesh generated by the proposed procedure, in which substantial element intersections were present.

---

[a] Use of the MSMS-default 1.5 Å radius probe resulted in QSLIM-decimated surface models that were poorly formed with multiple intersecting and degenerate triangles due to re-entrant surfaces of F-actin. Use of a 3 Å radius probe resolved this problem of SES-representation and is not expected to affect significantly the large length-scale normal modes of F-actin, which has relatively large minor and major diameters of ~40 and 80 Å, respectively.

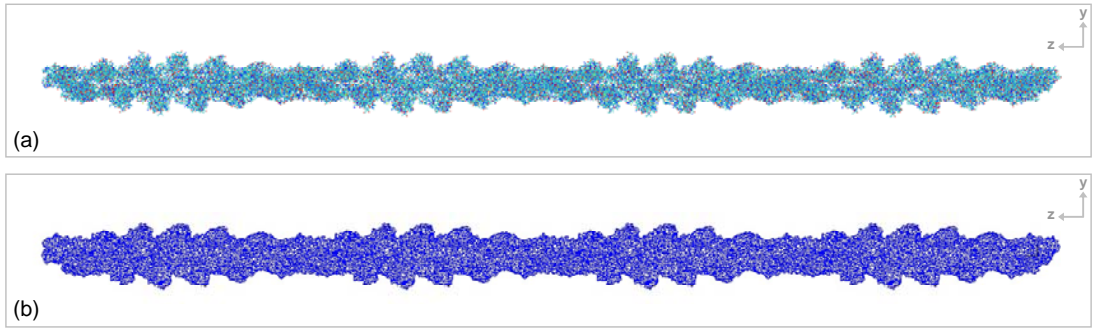

**Fig. 2** (a) Atomic structure of the 52-monomer F-actin filament analyzed and (b) the triangulated SES used to define the FE model. Atomic structure is rendered with VMD ver. 1.8.5 [59] and the FEM model rendered using ADINA ver. 8.4.

## Results
*T4 lysozyme*

Equilibrium thermal fluctuations of Cα atoms aid in understanding protein function as mediated by local conformational flexibility and provide a first quantitative test for the proposed coarse-grained procedure. Experimental fluctuations are related to the experimental temperature- or B-factor by, $B_i = 8\pi^2 \langle \Delta r_i^2 \rangle / 3$, where $\langle \Delta r_i^2 \rangle$ is the mean-squared fluctuation of atom *i*. While both coarse-grained models capture well the overall experimental variation in flexibility of T4 lysozyme (Fig. 3a and Table 1),[56] local differences are evident in disordered loop regions where conformational flexibility is overestimated significantly by both the RTB and FEM procedures (e.g., residue numbers 35–40). Comparison with the all-atom model indicates that these discrepancies are inherent to the protein structure, however, and not artifacts of the RTB and FEM procedures (Fig. 3b). Indeed, Cα fluctuations calculated with the RTB and FEM models correlate notably better with fluctuations calculated with the all-atom model than with experiment (Table 1).

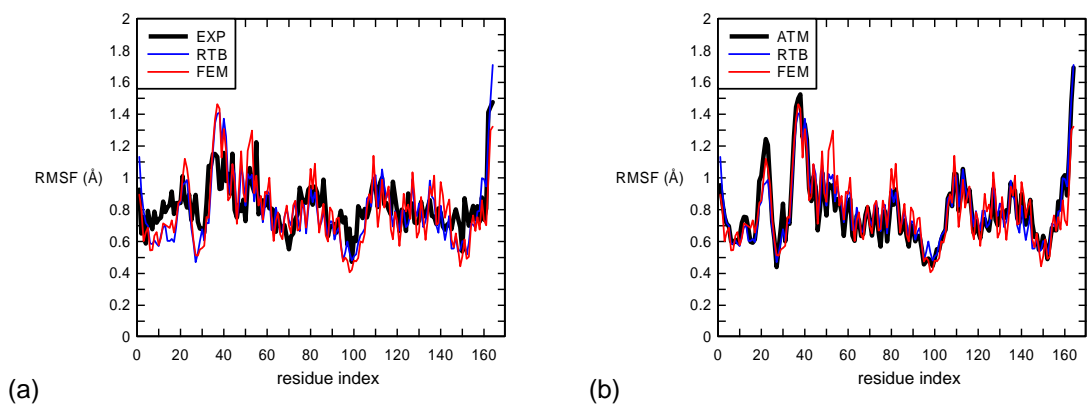

**Fig. 3.** Coarse-grained RMSF of Cα atoms in T4 lysozyme compared with (a) experiment and (b) all-atom NMA. 100 modes are used to compute the all-atom, RTB, and FEM fluctuations. Correlation coefficients provided in Table 1.

**Table 1** Correlation coefficients corresponding to Cα atom RMSF in Figure 3.

|  | **Experiment** | **ATM** |
|---|---|---|
| **RTB** | 0.73 | 0.95 |
| **FEM** | 0.68 | 0.89 |

Inter-residue spatial correlations measured at Cα atoms provide additional insight into protein function,[69,70] as well as a further test of the proposed coarse-grained procedure. Interestingly, the RTB and FEM procedures provide similar information with respect to the all-atom model, as measured over either the lowest 10 or 100 modes (Fig. 4 top and bottom, respectively). The fact that the correlation maps are largely determined with as few as ten modes reconfirms numerous previous findings that the lowest modes of proteins dominate their free vibration response.[14,71,72] The similarity in the FEM and ATM correlation maps provides additional evidence that T4 lysozyme behaves remarkably similar to a homogeneous isotropic elastic solid in free vibration.

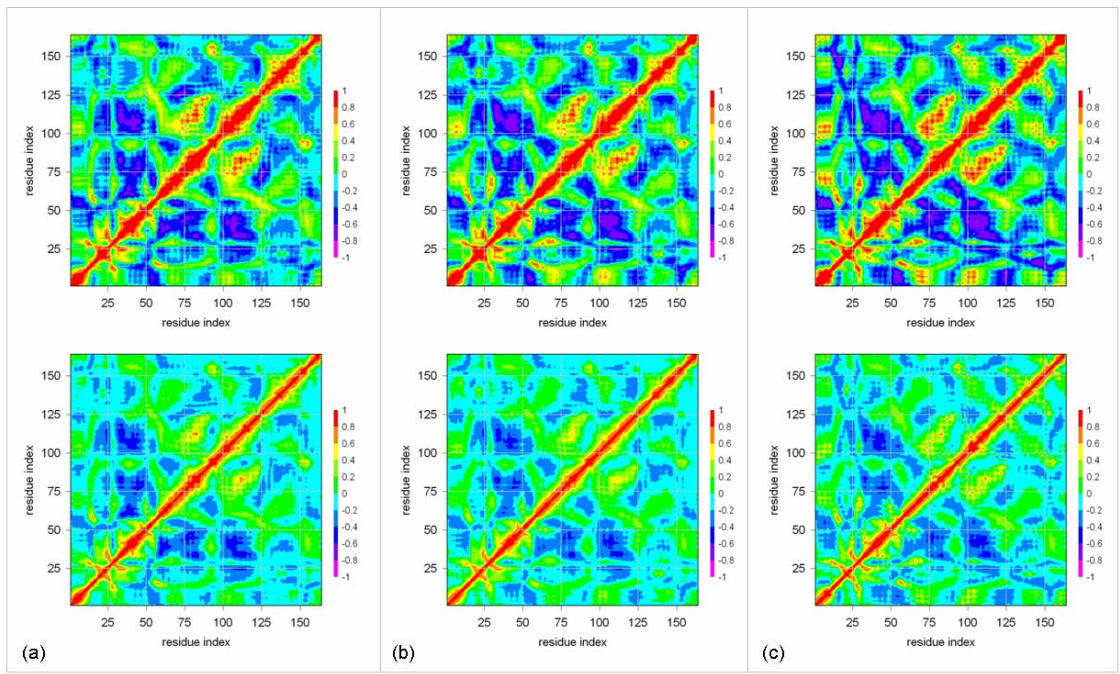

**Fig. 4** T4 lysozyme inter-Cα correlations computed using (top) 10 modes and (bottom) 100 modes for the (a) ATM, (b) RTB, and (c) FEM models.

The lowest four mode shapes computed using the FEM may be projected onto the ground-state (energy-minimized) structure of T4 lysozyme to visualize their nature (Fig. 5). Similar to the native hen egg lysozyme, the lowest mode is a hinge-bending mode,[1,73] whereas the three higher modes are a combination of hinge- and twist-deformations. Quantitative comparison between the coarse-grained and all-atom models is made in Table 2 for the lowest four mode shapes, and the lowest 200 frequencies in Figure 6.

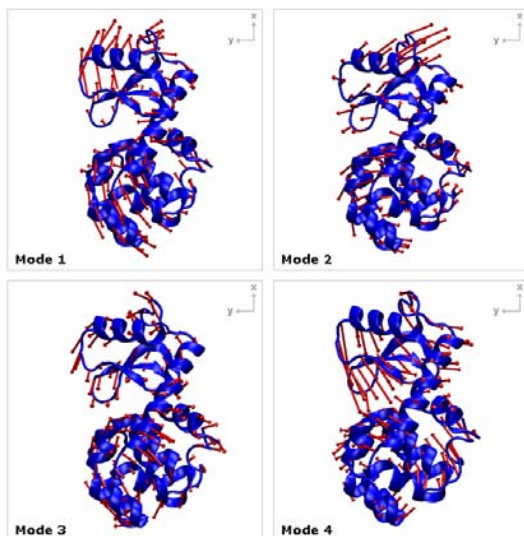

**Fig. 5** Lowest four eigenmodes computed by the FEM superimposed on the minimized structure of T4 lysozyme. Overlap with the all-atom model is given in Table 2. Images rendered using VMD ver. 1.8.5 [59].

**Table 2** Overlap of coarse-grained model and all-atom normal modes as measured at C$\alpha$ positions.

|  | **Mode 1** | **Mode 2** | **Mode 3** | **Mode 4** |
|---|---|---|---|---|
| **RTB** | 0.97 | 0.93 | 0.82 | 0.28 |
| **FEM** | 0.91 | 0.86 | 0.76 | 0.71 |

The modal frequency distributions provide a final quantitative evaluation of the FEM and RTB approach for T4 lysozyme (Fig. 6). While the overall correlation between the FEM and all-atom frequencies is reasonable, particularly for low mode-numbers, the FEM tends to *underestimate* the "exact" frequency computed using the all-atom model at high mode-numbers. This suggests that the FEM models the protein as overly compliant in this regime, which is to be expected because higher modes excite shorter wavelength, stiffer degrees of freedom in the all-atom protein resulting from chain connectivity, whereas the elastic solid approximation assumes a compliance that is length-scale invariant. Backbone C$\alpha$ fluctuations as well as C$\alpha$ correlations are apparently unaffected by this approximation because the low modes dominate these observables. Interestingly, the opposite tendency was observed by Tama et al.,[21] for the RTB-approach with successively larger blocks. This is also to be expected because the assumption of rigid blocks in the protein renders the structure overly stiff on short length scales (high frequency modes), and the length-scale at which this deviation from the all-atom model becomes significant increases with increasing block-size.

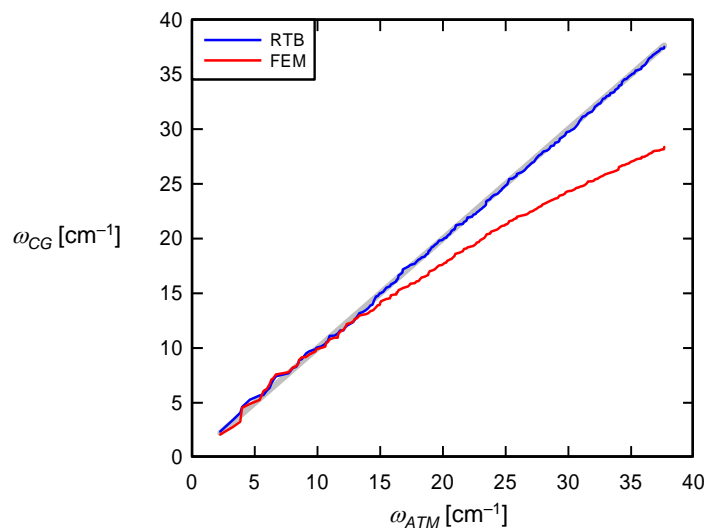

**Fig. 6** Correlation between coarse-grained (CG) and all-atom (ATM) model frequencies for the lowest 200 modes.

*F-actin*
F-actin is a highly dynamic biopolymer with a considerable degree of internal plasticity in the state of tilt and twist of its constituent protomers, which depends on the bound nucleotide-state (ATP/ADP), bound actin-binding protein, and solvent conditions.[74-78] Additionally, the bending stiffness of F-actin has been shown to increase by a factor of two in the presence of phalloidin, by 50% in the F-ADP-P versus F-ADP state, and to be regulated by tropomyosin in a $Ca^{2+}$-dependent fashion.[66] Thus, any modeling attempt to predict the mechanical properties of F-actin and investigate their relation to its detailed internal structure and composition must consider such variations.

Modeling attempts to investigate the structure-function relation of F-actin include an early study by ben-Avraham and Tirion,[68] who treated G-actin monomers as internally rigid and connected to their nearest neighbor monomers by compliant springs, a more recent study by Ming et al.,[79] in which conventional eNMA is used together with substructure-synthesis to calculate the large wavelength normal modes of a micron-long F-actin molecule, and most recently an all-atom MD study by Chu and Voth,[67] who found that the loop-helix transition of the DNase I binding region of subdomain 2 of G-actin plays a central role in respectively stabilizing-destabilizing F-actin by disrupting inter-monomer interactions. Chu and Voth[67] also calculated the apparent persistence length of F-actin and found that the loop-to-helix transition between the ATP- and ADP-bound states accounted for the approximately 50% decrease in associated bending stiffness observed experimentally.[66]

The normal modes of F-actin (52-mer, 0.14 $\mu$m length) computed here in free planar-vibration yield four bending modes as the lowest modes (Fig. 7). Association of F-actin with a homogeneous elastic rod in free vibration[80] results in an apparent bending stiffness, $\kappa = 6.8 \times 10^{-26}$ Nm$^2$, for the lowest mode, which is near the upper limit of bending stiffness typically reported experimentally.[66,81,82]

Subjecting F-actin to an axially compressive load and performing a linearized buckling analysis yields the lowest critical Euler buckling load, $P_{crit} = 33$ pN. Association of the filament again with a homogeneous elastic Euler–Bernoulli beam yields the effective bending stiffness, $\kappa = P_{crit} L^2 / \pi^2 = 6.9 \times 10^{-26}$ Nm$^2$, which is similar to the bending stiffness calculated from the lowest bending mode because that mode of deformation is the same as the lowest Euler buckling mode.

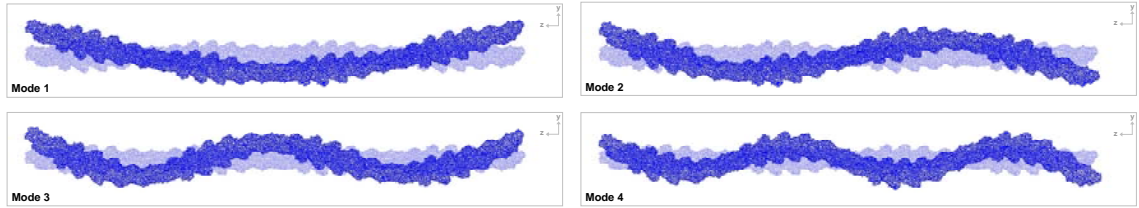

**Fig. 7** Four lowest free vibration modes of F-actin (52-mer, 0.144 $\mu$m length) in planar deformation. The corresponding angular frequencies are, $0.18\times10^{-2}$, $0.48\times10^{-2}$, $0.92\times10^{-2}$, and $0.16\times10^{-1}$ rad/psec.

The bending stiffness calculated here for F-actin is consistent with experimental measurements of the ATP-bound-state in which the DNase binding region (residues 40–48) in subdomain 2 of G-actin is in its disordered loop conformation, thereby stabilizing inter-monomer interactions.[66,67] While this is to be expected given the structure of G-actin:ADP:$Ca^{2+}$ employed, in which the DNase binding region is also in its disordered loop conformation, a similar coarse-grained analysis of F-actin must be performed in which the DNase binding region of G-actin is in its ordered α-helical structure. Only then may it be stated definitively whether the observed mechanical behavior is due solely to this detailed structural difference or to some other source, such as a lack of modeling resolution.

While a more detailed investigation of this type is of direct interest in evaluating the full utility of the proposed procedure, it is also of interest fundamentally to investigate the respective roles of molecular *shape* versus molecular *interactions* on determining the mechanical properties of supramolecular assemblies such as F-actin, MTs, and viral capsids. In particular, an intriguing hypothesis is that mechanical response is determined solely by molecular *shape*, in which case the mechanical properties of supramolecular assemblies would be robust to amino acid mutations that do not alter molecular shape. A competing hypothesis is that mechanical response is sensitive to both molecular shape and detailed molecular interactions, in which case amino acid mutations would be more tightly constrained. In either case, investigation of the respective roles of molecular shape versus specific interactions on protein mechanics clearly requires that all-atom models be considered, either directly or via incorporation into coarse-grained models. Such investigations are currently underway and are expected to provide fundamental insight into the origin and robustness of the mechanical properties of supramolecular assemblies.

**Concluding discussion**

A coarse-grained FE-based procedure is proposed to compute the normal modes and mechanical response of proteins and their supramolecular assemblies. The procedure takes as input the atomic structure to define uniquely the volume associated with the SES, mass density, and elastic stiffness of the protein. The initial, high resolution SES discretized at atomic resolution is simplified using a quadric simplification algorithm to obtain a molecular surface representation of arbitrary prescribed spatial resolution. While the proposed procedure is applied to proteins with known atomic structure, the molecular volume could equally be defined from electron density data, rendering the procedure applicable to a broad class of biomolecules and biomolecular complexes for which only a rough approximation to the molecular volume is known. As with existing coarse-grained elastic network models, energy minimization is not required prior to the NMA because the initial structure is assumed to be the ground-state structure.

Ongoing development of the proposed procedure is directed towards three areas of improvement. First, the atomic-based Hessian from all-atom force-fields such as CHARMM[58] will be projected onto the FE-space such that the model optimally converges to the "exact" all-atom solution as the FE mesh is refined to atomic length-scales. Such a procedure will enable the systematic coarsening of protein structure and interactions without the *a priori* assumption of elastic response. Indeed, an intriguing and as of yet unresolved question regards the relative effects of molecular *shape* versus specific molecular *interactions* on the mechanical response of supramolecular assemblies such as F-actin, MTs, and viral capsids. Second, the Poisson–Boltzmann equation used to model aqueous electrolyte-mediated electrostatic interactions in proteins may be coupled directly to the elastic-based FE model, so that it may be included in computations of normal modes and mechanical response. Langevin dynamics may also be incorporated into the model by coupling the protein-domain to the Stokes equations to model solvent damping.[41,83] Finally, the proposed surface discretization and simplification procedure requires improvement because it often results in surface meshes with intersecting or degenerate triangles, as encountered here for F-actin.

The utility of the proposed FE-based procedure is explored here for one specific globular protein and supramolecular assembly, namely T4 lysozyme and F-actin. Clearly, in order to evaluate the utility of the proposed procedure thoroughly, a set of proteins of drastically varying structure must be analyzed, as well as additional supramolecular assemblies. Additional response variables and the effects of internal structural variations of the molecules examined should also be investigated. Notwithstanding these additional analyses and the foregoing model improvements, the current communication establishes an effective theoretical framework for the computation of the normal modes and generalized mechanical response of proteins and their supramolecular assemblies based on the elastic medium theory of proteins.

**Acknowledgements**

Discussions with Marco Cecchini, Martin Karplus, Klaus–Jürgen Bathe, and Michael Garland are gratefully acknowledged, as is funding from the Alexander von Humboldt Foundation in the form of a post-doctoral fellowship. The author additionally thanks Michael Sanner for bringing QSLIM to his attention.